\documentclass[a4paper]{article}
\usepackage{pbox}
\usepackage{multirow}
\usepackage{cite}
\pdfoutput=1
\usepackage{INTERSPEECH2021}

\title{Speaker-conditioned Target Speaker Extraction \\ based on Customized LSTM Cells}
\name{Ragini Sinha$^1$, Marvin Tammen$^2$, Christian Rollwage$^1$, Simon Doclo$^{1,2}$}
\address{
  $^1$Fraunhofer Institute for Digital Media Technology, Project group Hearing, Speech and Audio Technology, Oldenburg, Germany\\
  $^2$Dept. of Medical Physics and Acoustics and Cluster of Excellence
Hearing4all, University of Oldenburg, Germany}
\email{ragini.sinha@idmt.fraunhofer.de, marvin.tammen@uni-oldenburg.de, christian.rollwage@idmt.fraunhofer.de, simon.doclo@uni-oldenburg.de}

\begin{document}

\maketitle

\begin{abstract}
Speaker-conditioned target speaker extraction systems rely on auxiliary information about the target speaker to extract the target speaker signal from a mixture of multiple speakers. Typically, a deep neural network is applied to isolate the relevant target speaker characteristics. In this paper, we focus on a single-channel target speaker extraction system based on a CNN-LSTM separator network and a speaker embedder network requiring reference speech of the target speaker. In the LSTM layer of the separator network, we propose to customize the LSTM cells in order to only remember the specific voice patterns corresponding to the target speaker by modifying the information processing in the forget gate. Experimental results for two-speaker mixtures using the Librispeech dataset show that this customization significantly improves the target speaker extraction performance compared to using standard LSTM cells.

\end{abstract}


\section{Introduction}
Recently, the problem of speaker extraction has attracted a lot of attention in the speech processing community. The goal of speaker extraction is to extract a target speaker signal from a mixture of multiple speakers. In principle, this can be achieved by first extracting all individual speakers from the mixture using blind source separation \cite{makino2018audio,vincent2018audio, hershey2016deep,kolbaek2017multitalker,luo2019conv,le2019phasebook} and then selecting the extracted signal that corresponds to the target speaker. However, a disadvantage of such a two-step approach is that the number of speakers needs to be known or estimated, which is not trivial in practice. Alternatively, one-step approaches have been proposed which aim at directly extracting the target speaker from the mixture by utilizing auxiliary information about the target speaker \cite{ vzmolikova2019speakerbeam, wang2018voicefilter, li2020atss, ge2020spex+, zhang2020x, ephrat2018looking, afouras2018conversation, li2020listen, gu2019neural, delcroix2021speaker}. Such techniques are also referred as speaker-conditioned target speaker extraction. Commonly used auxiliary information is reference speech of the target speaker \cite{vzmolikova2019speakerbeam, wang2018voicefilter, li2020atss, ge2020spex+, zhang2020x}, video of the target speaker \cite{ephrat2018looking, afouras2018conversation, li2020listen, michelsanti2020overview}, directional information \cite{gu2019neural,brendel2020unified,aroudi2020dbnet}, or information about the speech activity of the target speaker \cite{delcroix2021speaker}. In this paper, we focus on single-channel target speaker extraction using reference speech as auxiliary information.

The single-channel speaker-conditioned target speaker extraction systems in \cite{wang2018voicefilter, li2020atss, ge2020spex+, zhang2020x} consist of two networks: a separator network and a speaker embedder network. The speaker embedder network is used to generate embeddings from the reference speech of the target speaker. These embeddings are used along with the mixture in the separator network with the goal to extract the target speaker signal from the mixture. The separator networks in \cite{wang2018voicefilter, li2020atss} estimate a time-frequency mask to perform speaker extraction, whereas the separator networks in \cite{ge2020spex+, zhang2020x} perform speaker extraction in the time-domain. The embedder networks in \cite{wang2018voicefilter, zhang2020x} utilize the same LSTM-based architecture of a speaker verification system proposed in \cite{wan2018generalized}, whereas different ResNet-based speaker verification systems are utilized in \cite{li2020atss, ge2020spex+} to generate embeddings of the target speaker. In \cite{ge2020spex+} the speaker embedder network and the separator network are trained jointly using a weighted combination of a cross-entropy loss function and a scale-invariant signal-to-noise ratio (SI-SNR) loss function, while in \cite{wang2018voicefilter, li2020atss, zhang2020x} the speaker embedder network and the separator network are trained separately. The system in \cite{wang2018voicefilter} utilizes a convolutional long short-term memory (CNN-LSTM) separator network trained with a power-law compression loss function, while the system in \cite{li2020atss} utilizes an attention-based separator network trained with a squared l2-norm loss function and the system in \cite{zhang2020x} utilizes a separator network similar to \cite{luo2019conv} trained with an SI-SNR loss function. In this paper, our proposed speaker-conditioned target speaker extraction system is mostly inspired by the system presented in \cite{wang2018voicefilter} (VoiceFilter).

In order to improve the performance of the baseline system in \cite{wang2018voicefilter}, in this paper we propose two modifications. First, instead of using standard LSTM cells in the separator network we introduce an LSTM cell that is customized for speaker-conditioned target speaker extraction. The purpose of this customization is to force the forget gate of the LSTM cell to only remember the specific voice patterns corresponding to the target speaker, which are needed to distinguish between the target speaker and the other speakers present in the mixture. Second, instead of using the spectral-domain power-law compression loss function we use the time-domain SI-SNR loss function. Experimental results for two-speaker mixtures from the Librispeech dataset show that the target speaker extraction performance can be significantly improved compared to the baseline system \cite{wang2018voicefilter} in terms of signal-to-distortion ratio (SDR) \cite{vincent2006performance} and perceptual evaluation of speech quality (PESQ) \cite{itu-t_perceptual_2001}.

The remainder of this paper is organized as follows. In Section \ref{method} we introduce the considered target speaker extraction system, where we mainly focus on the proposed customization of the LSTM cells. Section \ref{exp} discusses the network architectures and the parameters used for training and evaluation. Section \ref{result} presents the target speaker extraction results.

\section{Target Speaker Extraction System} \label{method}
We consider a scenario where a single microphone records the mixture of $I$ speakers, i.e., in the time-domain the microphone signal can be written as
\begin{equation}
  y(n) = \sum_{i=1}^{I} x_i(n),
  \label{eq1}
\end{equation}
where $x_i(n)$ denotes the speech signal of the $i$-th speaker and $n$ denotes the discrete-time index. The goal is to extract the speech signal corresponding to the target speaker from the mixture, where without loss of generality the $j$-th speaker will be assumed to be the target speaker. 

\begin{figure}[th]
  \centering
  \includegraphics[width=\linewidth]{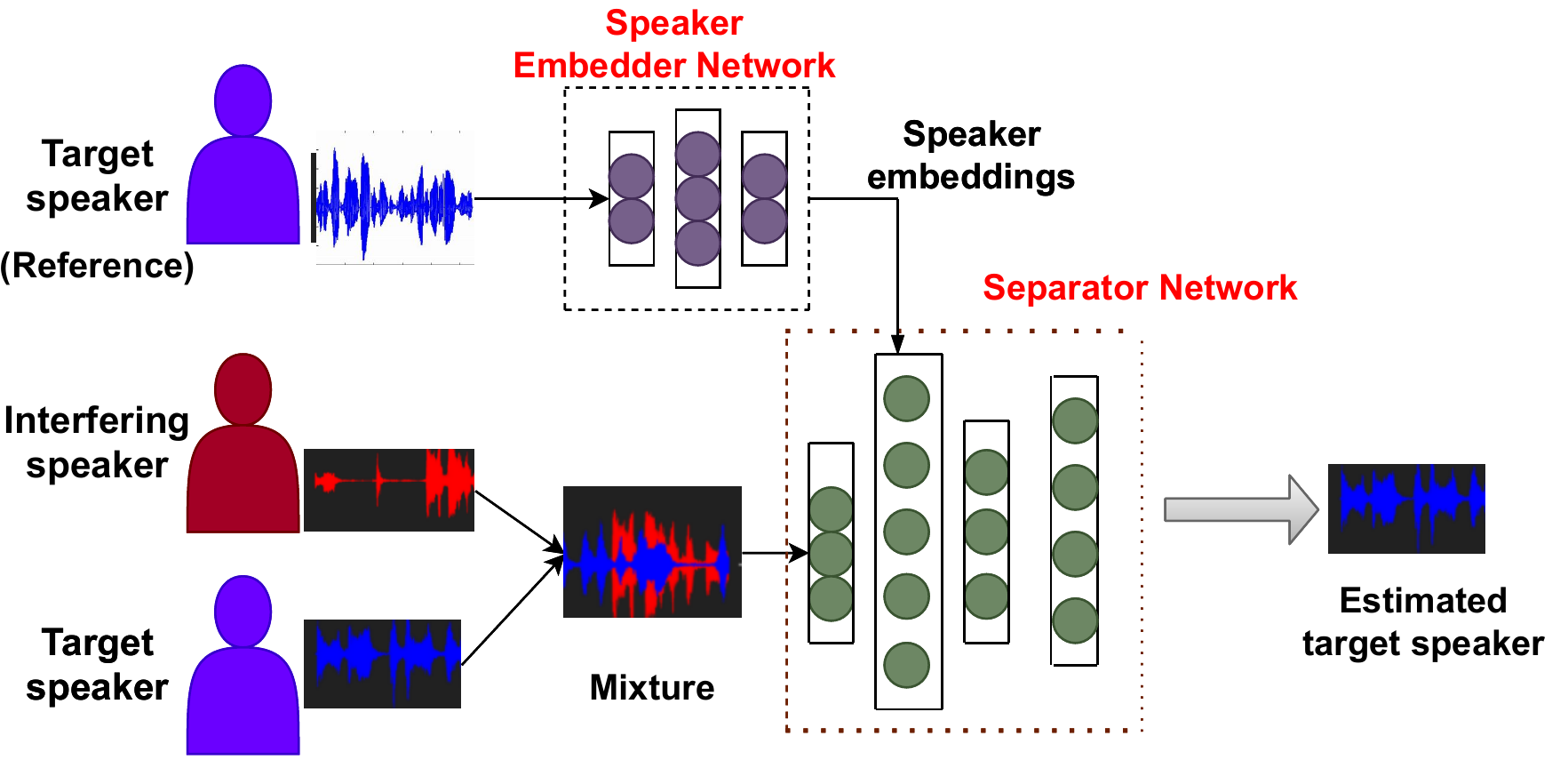}
  \caption{Block diagram of the used speaker-conditioned target speaker extraction system, consisting of a speaker embedder network and a separator network.}
  \label{fig:speaker-conditioning-training}
\end{figure}

Similarly to \cite{wang2018voicefilter, li2020atss, ge2020spex+, zhang2020x}, we consider a speaker-conditioned target speaker extraction system consisting of two parts (see Figure \ref{fig:speaker-conditioning-training}), namely, a speaker embedder network and a separator network. The speaker embedder network is used to generate the embeddings from the reference speech of the target speaker. Based on the target speaker embeddings and the mixture, the separator network aims at extracting the target speaker signal from the mixture. In this section, we will mainly discuss the separator network, while details about the embedder network will be provided in Section \ref{embedder}.

In \cite{wang2018voicefilter, li2020atss} speaker extraction was performed in the short-time Fourier transform (STFT) domain. The magnitude spectrum of the target speaker signal was estimated from the magnitude spectrum of the microphone signal using a (soft) mask $M(k,l)$, i.e.,
\begin{equation}
 \tilde{X}_j(k, l) = {M(k, l)}{Y(k, l)}, 
  \label{eq2}
\end{equation}
where $k$ and $l$ denote the frequency index and the time frame index, respectively. Using the CNN-LSTM network proposed in \cite{wang2018voicefilter}, the mask was computed from the target speaker embeddings $\mathbf{e}_j$ and the mixture as
\begin{equation}
  M(k, l) = \boldsymbol{\phi}{(\mathbf{r}, \mathbf{e}_j)}, 
  \label{eq3}
\end{equation}
\begin{equation}
 \mathbf{r} = \mathbf{g}(\lvert{Y(k, l)}\rvert), 
  \label{eq4}
\end{equation}
where $\mathbf{g}(\circ)$ denotes the convolutional layers of the separator network used to obtain the intermediate representation $\mathbf{r}$, and $\boldsymbol{\phi}(\circ)$ denotes the rest of the separator network. 

Instead of using standard LSTM cells in $\boldsymbol{\phi}(\circ)$, we propose to customize the information processing through the forget gate of the LSTM cells such that the specific voice patterns of only the target speaker are remembered, which helps to distinguish the target speaker from the other speakers in the mixture when performing the extraction. After briefly reviewing the standard LSTM cell used in the baseline system \cite{wang2018voicefilter} in Section \ref{general LSTM-cell}, in Section \ref{custom-lstm} we discuss the proposed customization of the LSTM cell.

\subsection{Standard LSTM cell} \label{general LSTM-cell}
\begin{figure}[th]
  \centering
  \includegraphics[width=\linewidth]{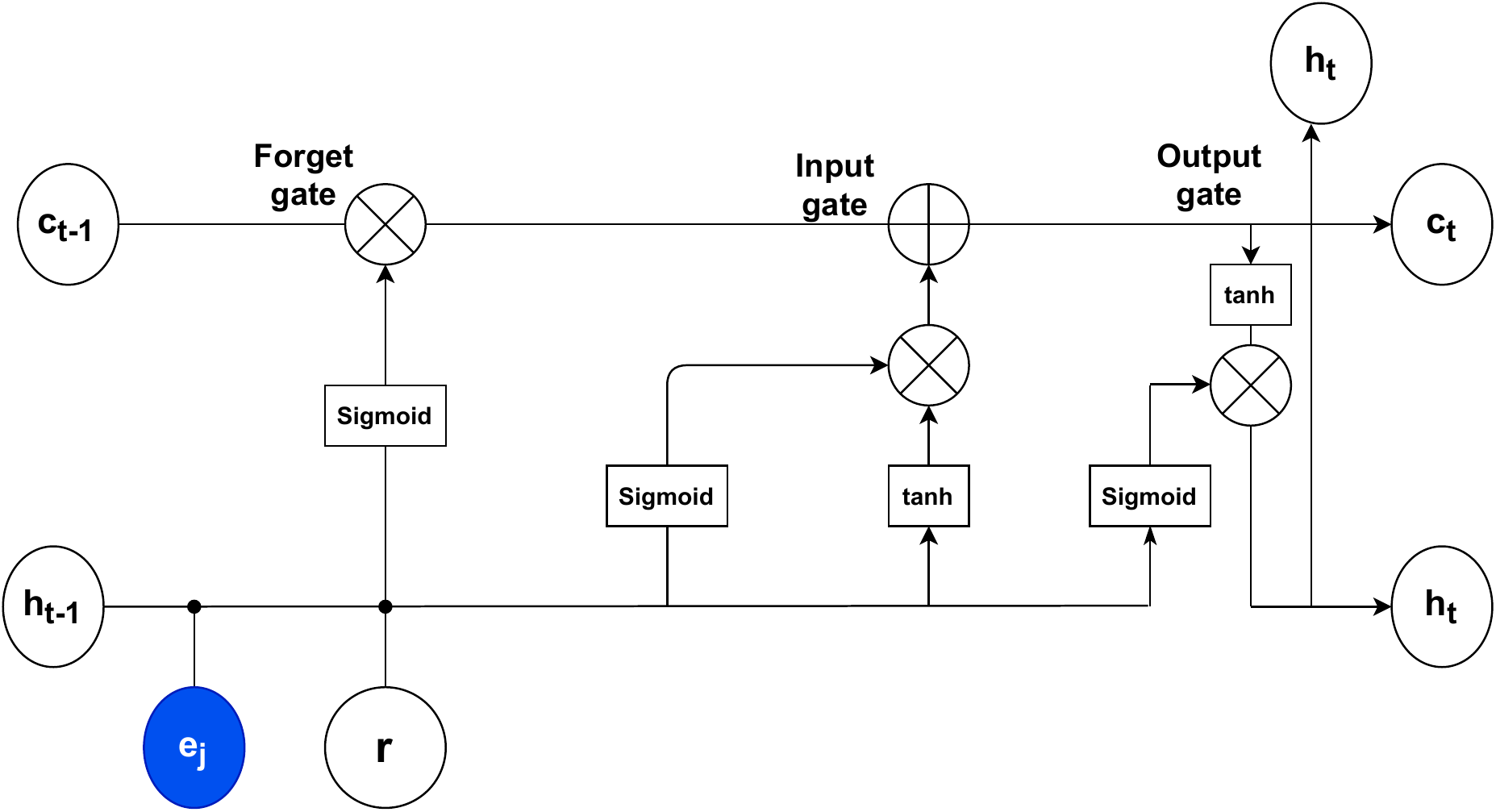}
  \caption{Standard LSTM cell.}
  \label{fig:LSTM-cell}
\end{figure}

The working principle of an LSTM cell \cite{hochreiter1997long, sherstinsky2020fundamentals} is determined by its state and three gates: the forget gate, the input gate, and the output gate (see Figure \ref{fig:LSTM-cell}). The cell state behaves like the memory of the network with the ability to retain information through time, while the gates can add or remove information at each step $t$. In the following, the weight matrix and the bias of the forget gate, the input gate, the output gate and the control update are denoted by $\mathbf{W}_{f}$, $\mathbf{W}_i$, $\mathbf{W}_c$, $\mathbf{W}_o$ and $\mathbf{b}_f$, $\mathbf{b}_i$, $\mathbf{b}_c$, $\mathbf{b}_o$, respectively. The current and previous cell states are denoted by $\mathbf{c}_{t}$ and $\mathbf{c}_{t-1}$, while the current and previous hidden states are denoted by $\mathbf{h}_{t}$, $\mathbf{h}_{t-1}$. 

As can be seen from (\ref{eq3}), the input to the LSTM layer (and hence each LSTM cell) is the concatenation of the speaker embeddings ${\mathbf{e}_j}$ and the output of the convolutional layers $\mathbf{r}$. The recursive nature allows the LSTM cell to store information from the previous state. The forget gate decides which information should be retained or disregarded based on the previous hidden state and the current input. The information is retained in the cell state if the output of the forget gate is close to 1, otherwise it is disregarded. The output of the forget gate is obtained as
\begin{equation}
  \mathbf{f}_t = \sigma(\mathbf{W}_f[\mathbf{h}_{t-1}, (\mathbf{r}, \mathbf{e}_j)] + \mathbf{b}_f),
  \label{eq5}
\end{equation}
where $\sigma$ denotes the Sigmoid activation function. 

The input gate decides which information is updated and stored in the cell state, and its output is obtained as
\begin{equation}
  \mathbf{i}_t = \sigma(\mathbf{W}_i[\mathbf{h}_{t-1}, (\mathbf{r}, \mathbf{e}_j)] + \mathbf{b}_i).
  \label{eq6}
\end{equation}
The cell state behaves like the memory of the network, and is updated as
\begin{equation}
  \mathbf{\tilde{c}}_t = \tanh(\mathbf{W}_c[\mathbf{h}_{t-1}, (\mathbf{r}, \mathbf{e}_j)] + \mathbf{b}_c),
  \label{eq7}
\end{equation}
\begin{equation}
  \mathbf{c}_t = \mathbf{f}_t \ast \mathbf{c}_{t-1} + \mathbf{i}_t \ast {\mathbf{\tilde{c}}_t},
  \label{eq8}
\end{equation}
where $\ast$ denotes point-wise multiplication.
Finally, the output gate decides which part of the cell state is transferred to the next hidden state, and its output is obtained as
\begin{equation}
  \mathbf{o}_t = \sigma(\mathbf{W}_o[\mathbf{h}_{t-1}, (\mathbf{r}, \mathbf{e}_j)] + \mathbf{b}_o).
  \label{eq9}
\end{equation}
Finally, the hidden state is updated as
\begin{equation}
  \mathbf{h}_t = \mathbf{o}_t \ast \tanh{(\mathbf{c}_t)}.
  \label{eq10}
\end{equation}
 
\subsection{Customized LSTM cell} \label{custom-lstm}
\begin{figure}[th!]
  \centering
  \includegraphics[width=\linewidth]{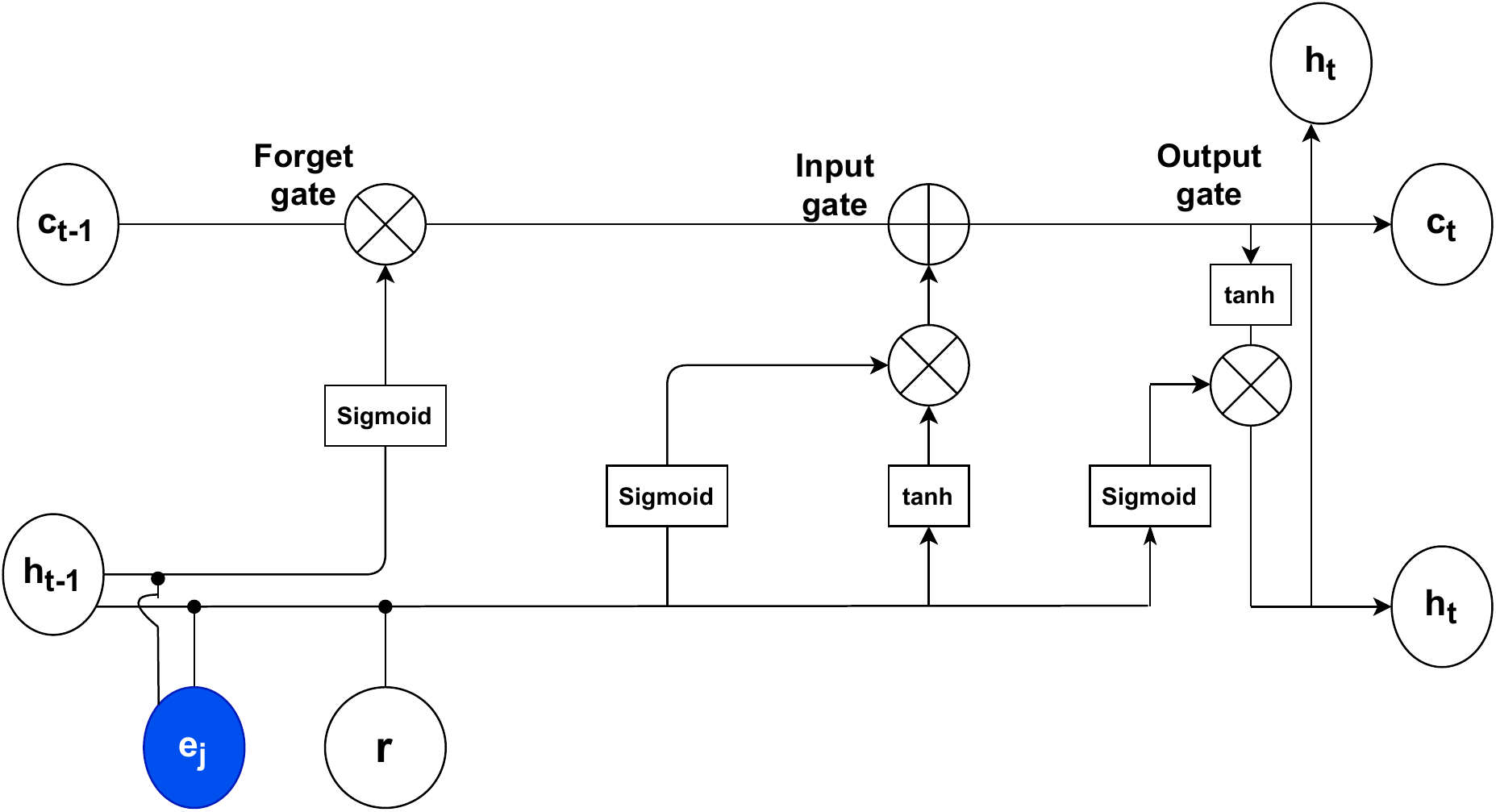}
    \caption{Proposed customized LSTM cell for speaker-conditioned 
    target speaker extraction.}
    \label{fig:cutomized-LSTM-cell}
\end{figure}

As already mentioned, the forget gate is used to retain the relevant information and disregard irrelevant information, based on the previous hidden state and the current input. With the specific goal of speaker extraction in mind, intuitively the LSTM cell is supposed to learn to retain information related to the target speaker, while disregarding information unrelated to the target speaker, i.e.,  originating from the other speakers present in the mixture. However, since in practice this will not be perfectly achieved, we propose to customize the LSTM cell in order to only retain the target speaker information by changing the information processing through the forget gate (see Figure \ref{fig:cutomized-LSTM-cell}). Instead of considering the concatenation of the target speaker embeddings $\mathbf{e}_j$ and the output of the convolutional layer $\mathbf{r}$, we only consider the target speaker embeddings, i.e., 
\begin{equation}
  \mathbf{f}_t = \sigma(\mathbf{W}_e[\mathbf{h}_{t-1}, \mathbf{e}_j] + \mathbf{b}_e),
  \label{eq11}
\end{equation}
where $\mathbf{W}_e$ and $\mathbf{b}_e$ denote the weight matrix and the bias of the customized forget gate. It should be noted that all other gates, i.e., the input and output gates, and the cell update remain the same as described in the previous section. 
 
In the customized LSTM cell, the forget gate in (\ref{eq11}) aims at mapping the target speaker close to 1. This allows the current cell state in (\ref{eq8}) to retain the target speaker information by multiplying the previous cell state with a value close to 1, while disregarding the information related to the other speakers from the previous cell state $\mathbf{c}_{t-1}$. We only customize the forget gate, since the forget gate is the main gate modifying the cell state. We do not customize the input and output gates. Since the input gate can only add but cannot remove information from the current cell state, a similar customization for the input gate might not be as effective as for the forget gate. A similar customization for the output gate may even lead to the loss of relevant information in the next hidden state.

\section{Experimental Setup} \label{exp}
In this section, we discuss the network architecture, the used parameters and the training procedure for the speaker embedder network and the separator network.

\subsection{Speaker Embedder Network} \label{embedder}
We have used the same speaker embedder network as the baseline system \cite{wang2018voicefilter}, namely the speaker verification network originally proposed in \cite{wan2018generalized}. It consists of 3 LSTM layers, each having 768 nodes. As input features the network uses 40-dimensional log-Mel-features, which are computed using an STFT size of 512 at a sampling frequency of 16 kHz, a Hanning window with a frame length of 400 and a frame shift of 160 samples. Given the log-Mel-features of the reference speech of a target speaker, the speaker embedder network generates 256-dimensional embeddings to be utilized further in the separator network. 
We have used the official training and validation split of the Voxceleb dataset \cite{nagrani2017voxceleb} to retrain the speaker embedder network. The Voxceleb dataset is a large speaker verification dataset that consists of more than one million utterances from more than 7000 speakers.

\subsection{Separator Network} \label{separator}
Similarly as for the baseline system \cite{wang2018voicefilter}, the separator network consists of eight 2D dilated convolutional layers, a customized LSTM layer and two fully connected (FC) layers. Each convolutional layer is followed with a batch-normalization layer and a ReLU activation function, where the dilated convolutional layers are used to increase the size of the receptive field of the network. The number of filters, filter-size and the dilation parameters for the convolutional layers are the same as for the baseline system \cite{wang2018voicefilter} (see Table \ref{tab:parameters}). The only difference in the architecture of the proposed separator network compared to the baseline is the number of nodes for the customized LSTM layer and the FC layers. We have used 600 nodes for the customized LSTM layer and the first FC layer consists of 514 nodes followed with a ReLU activation function, while the second FC layer consists of 257 nodes with a Sigmoid activation function. As input features the network uses the magnitude of the STFT coefficients of the mixture. The STFT coefficients are computed using an STFT size of 512 at a sampling frequency of 16 kHz, a square-root Hanning window with a frame length of 512 and a frame shift of 256 samples. 

To train the separator network we have considered two different loss functions: a spectral-domain loss function based on power-law compression (PLC) \cite{wang2018voicefilter}, and the scale-invariant signal-to-noise ratio (SI-SNR) loss function in the time-domain \cite{luo2019conv}. In total we have trained three different separator networks:
\begin{enumerate}
  \item \textbf{Standard / PLC}: using the PLC loss function and the standard LSTM cells. This separator network is similar to the baseline system in \cite{wang2018voicefilter}. Since the baseline system is not publicly available, we retrained it using the same loss function and model parameters.
  \item \textbf{Standard / SI-SNR}: using the SI-SNR loss function and standard LSTM cells.
  \item \textbf{Customized / SI-SNR}: using the SI-SNR loss function and the proposed customized LSTM cells.
\end{enumerate}
The Adam optimizer \cite{kingma2014adam} with a learning rate of 0.0002 was used. We used a batch size of 16 and fixed the total number of epochs to 50, while clipping the gradient norm to 10. An early stopping criterion was used to prevent the network from further training if the validation loss did not decrease after 7 epochs.

\begin{table}[th]
  \caption{Parameters of the proposed separator network.}
  \label{tab:parameters}
  \centering
  \begin{tabular}{|l|c|c|c|} 
    \hline
    Layer & Kernel Size & Dilation & Filters/Nodes \\ [1ex] 
    \hline\hline
    Conv1 & $(1\times7)$ & $(1\times1)$ & 64 \\ 
    \hline
    Conv2 & $(7\times1)$ & $(1\times1)$ & 64 \\
    \hline
    Conv3 & $(5\times5)$ & $(2^{0}\times1)$ & 64 \\
    \hline
    Conv4 & $(5\times5)$ & $(2^{1}\times1)$ & 64 \\
    \hline
    Conv5 & $(5\times5)$ & $(2^{2}\times1)$ & 64 \\
    \hline
    Conv6 & $(5\times5)$ & $(2^{3}\times1)$ & 64 \\ 
    \hline
    Conv7 & $(5\times5)$ & $(2^{4}\times1)$ & 64 \\ 
    \hline
    Conv8 & $(1\times1)$ & $(1\times1)$ & 8 \\
    \hline
    \pbox{2cm}{Customized\\LSTM} & -& -& 600 \\[1.5ex] 
    \hline
    FC 1 & - &- &514 \\
    \hline 
    FC 2 &- & -& 257 \\[1ex] 
    \hline
\end{tabular}
\end{table}

To generate the training data for our separator networks, we have used 100 hours of clean speech from the Librispeech dataset \cite{panayotov2015librispeech}. We have followed the same procedure as in \cite{wang2018voicefilter} to create the mixture, the reference speech and the target speech, each having a length of 4 seconds. The mixture is constructed (assuming $I$=2 speakers) by summing the utterances of two randomly chosen different speakers (target and interfering) from the dataset, while the reference speech is a randomly chosen utterance of the target speaker, which is completely different from the utterances used for constructing the mixture.

\section{Results and Discussion} \label{result}
We have evaluated the performance of all considered target speaker extraction systems using the official test set of the Librispeech dataset. The same procedure as for training has been used to generate the mixture, the reference speech and the target speech (see Section \ref{separator}). As performance measures, we have used the signal to distortion ratio (SDR) \cite{vincent2006performance} and the perceptual evaluation of speech quality (PESQ) \cite{itu-t_perceptual_2001} measure. SDR is a common measure to evaluate the performance of source separation systems, while PESQ is a speech quality metric commonly used for speech enhancement systems. For both performance measures, the clean target speech signal was used as the reference signal. 

For the three considered target speaker extraction systems, Table \ref{tab:results-SDR} shows the mean SDR improvement (in dB) and the mean PESQ improvement with respect to the mixture. For reference, this table also shows the mean SDR improvement of the baseline (VoiceFilter) reported in \cite{wang2018voicefilter}. We do not show the mean PESQ improvement for the baseline as no PESQ results were reported in \cite{wang2018voicefilter}. For the mixture, the mean SDR is equal to 0.14 dB and the mean PESQ is equal to 1.37. First, it can be observed that the mean SDR improvement of our retrained system using standard LSTM cells and PLC loss function is very similar to the mean SDR improvement of the original VoiceFilter \cite{wang2018voicefilter}, showing the validity of retraining the networks. Second, it can be observed that by using the SI-SNR loss function instead of the PLC loss function (with standard LSTM cells), the mean SDR is improved by 1.37 dB, while the mean PESQ is improved by 0.06. Third, by using the proposed customized LSTM cells (with the SI-SNR loss function) the mean SDR can be further improved by 0.97 dB, while the mean PESQ can be further improved by 0.06. These results show that both proposed modifications significantly improve the performance compared to the baseline target speaker extraction system. 

\begin{table}[th]
  \caption{Mean SDR improvement ($\Delta{}$SDR) and mean PESQ improvement ($\Delta{}$PESQ) of baseline system (VoiceFilter), retrained system using standard LSTM cells and PLC loss function, system using standard LSTM cells and SI-SNR loss function, and proposed system using customized LSTM cells and SI-SNR loss function.}
  \label{tab:results-SDR}
  \centering
  \begin{tabular}{|l|c |c|} 
    \hline
    \pbox{2.8cm}{Target Speaker \\ Extraction System} & \pbox{2.8cm}{Mean \\ $\Delta{}$SDR (dB)} & \pbox{2.8cm}{Mean \\ $\Delta{}$PESQ} \\ [2ex] 
    \hline\hline
    Baseline (VoiceFilter) \cite{wang2018voicefilter} & 5.50 &- \\ 
    \hline\hline
    Standard / PLC & 5.62 & 1.95 \\
    \hline
    Standard / SI-SNR & 6.99 & 2.01 \\
    \hline
    Customized / SI-SNR & \textbf{7.96} & \textbf{2.07} \\[1ex] 
    \hline
\end{tabular}
\end{table}

\section{Conclusion} \label{conclusion}
For a CNN-LSTM target speaker extraction system, in this paper we have proposed customized LSTM cells aimed at retaining information of the target speaker and disregarding irrelevant information of the other speakers. To this end, we have modified the forget gate of the LSTM cell, considering only the target speaker embeddings, while keeping the other gates and the cell state update unchanged. Experimental results for two-speaker mixtures using the Librispeech dataset show that the proposed customization yields performance improvements in terms of both SDR and PESQ compared to using standard LSTM cells. In future work, we will investigate the usage of the proposed customized LSTM cells when jointly training of the speaker embedder network and the separator network, and for other auxiliary information of the target speaker.

\bibliographystyle{IEEEtran}
\nocite{*}
\bibliography{main}


\end{document}